\documentclass[aps,prx,superscriptaddress,amsmath,amssymbs, twocolumn]{revtex4-2}

\usepackage{xcolor}
\usepackage{graphicx}
\usepackage{braket}
\usepackage{float}
\usepackage{units} 
\usepackage{soul} 
\usepackage{verbatim} 	
\usepackage[hidelinks]{hyperref}
\usepackage{bm}
\usepackage{SIunits}
\usepackage{bigints}
\usepackage{svg}
\begin{document}

\title{Bandwidth enhanced current noise suppression with a SERF magnetometer}

\author{Tobias Menold}
\email{tobias.menold@ifsw.uni-stuttgart.de}
\affiliation{Center for Quantum Science, Physikalisches Institut, 
Eberhard Karls Universit\"at T\"ubingen, Auf der Morgenstelle 14, 
D-72076 T\"ubingen, Germany}
\affiliation{Institut f\"ur Strahlwerkzeuge (IFSW), Universit\"at Stuttgart, Pfaffenwaldring 43,
D-70596 Stuttgart, Germany}
\author{Arianna Bertoluzza}
\affiliation{Center for Quantum Science, Physikalisches Institut, 
Eberhard Karls Universit\"at T\"ubingen, Auf der Morgenstelle 14, 
D-72076 T\"ubingen, Germany}
\author{Patrick Hildebrand}
\affiliation{Center for Quantum Science, Physikalisches Institut, 
Eberhard Karls Universit\"at T\"ubingen, Auf der Morgenstelle 14, 
D-72076 T\"ubingen, Germany}
\affiliation{Institut f\"ur Strahlwerkzeuge (IFSW), Universit\"at Stuttgart, Pfaffenwaldring 43,
D-70596 Stuttgart, Germany}
\author{Ann-Kathrin Gottschalk}
\affiliation{Center for Quantum Science, Physikalisches Institut, 
Eberhard Karls Universit\"at T\"ubingen, Auf der Morgenstelle 14, 
D-72076 T\"ubingen, Germany}
\author{Daniel Braun}
\affiliation{Center for Quantum Science, Physikalisches Institut, 
Eberhard Karls Universit\"at T\"ubingen, Auf der Morgenstelle 14, 
D-72076 T\"ubingen, Germany}
\author{J\'{o}zsef Fort\'{a}gh}
\affiliation{Center for Quantum Science, Physikalisches Institut, 
Eberhard Karls Universit\"at T\"ubingen, Auf der Morgenstelle 14, 
D-72076 T\"ubingen, Germany}
\author{Andreas G\"unther}
\email{a.guenther@uni-tuebingen.de}
\affiliation{Center for Quantum Science, Physikalisches Institut, 
Eberhard Karls Universit\"at T\"ubingen, Auf der Morgenstelle 14, 
D-72076 T\"ubingen, Germany}

\begin{abstract}
\noindent
This work investigates the behavior of a spin-exchange relaxation-free (SERF) magnetometer integrated into the feedback branch of a closed-loop control circuit, designed to actively suppress noise from a current source. In this configuration, the magnetometer bandwidth is enhanced by almost two orders of magnitude as compared to the open-loop setup. Incorporating an injection transformer, the proposed system effectively reduces current noise to below $\unit{3}{\nano\ampere/\sqrt{\hertz}}$ over a broad frequency bandwidth, independent of DC-current offsets. Analysis of the system sensitivity reveals that the performance is currently limited by technical noise. If this were to be eliminated, the system would achieve a current sensing sensitivity on the order of few$\unit{}{\pico\ampere/\sqrt{Hz}}$, while maintaining perfect galvanic isolation from the target circuit.

\end{abstract} 

\date{\today}

\maketitle

\noindent

\section{Introduction}
\label{sec:introduction}

Spin-exchange relaxation-free (SERF) optical magnetometers are among the most sensitive magnetic field sensors currently available. They are based on pumping and probing the spin polarization of atomic ensembles in thermal vapour cells and can achieve sensitivities in the sub-fT/$\sqrt{\text{Hz}}$ range  
\cite{allred2002high, kominis2003subfemtotesla, ledbetter2008spin}. 
One of their most promising applications lies in the detection of neuronal activities, 
with potential uses in medical diagnostics and human-machine interfaces 
\cite{li2018serf, boto2017new, zheng2024compact, yang2021new, su2024vector, brookes2022magnetoencephalography}. 
Additional areas under exploration include material testing and fundamental physics research 
\cite{koss2022optically, thiemann2022using}.

Recent studies have shown promising results when SERF magnetometers are integrated into the feedback branch of a closed-loop control circuit designed to suppress current noise 
\cite{shen2020suppression, koss2021optical, mrozowski2023ultra}. These results are particularly encouraging in the sub-Hz frequency regime, where SERF magnetometers appear well suited for current noise detection. This changes in the higher frequency domain, as SERF magnetometers typically operate at a limited dynamic range, with a cut-off frequency roughly on the order of $\unit{100}{Hz}$. Despite this limitation, some studies suggest that integrating SERF magnetometers into closed-loop systems could help to extend their operational bandwidth \cite{zheng2024compact, guo2021high, jia2024comprehensive, liu2024performance}. However, a detailed study, especially in the context of current noise suppression, is still pending.

In this combined theoretical and experimental study, we address the fundamental question of how exactly a SERF magnetometer behaves in a closed-loop circuitry and how this can be used to increase the bandwidth of the system and realize efficient noise suppression for conventional current sources. Our studies are performed with a cesium (Cs) magnetometer based on an atomic vapour cell and a corresponding pump-probe laser scheme.
We show that the SERF magnetometer behaves like a first-order low-pass filter and that a feedback control system can be used to increase its bandwidth. As compared to the open-loop scenario, we experimentally demonstrate a 66-fold bandwidth improvement, which we use to realize efficient current noise suppression. We present an electronic circuitry, based on an injection transformer, which allows to reduce the current noise of a commercial voltage-controlled current source to about $\unit{2.5}{\nano\ampere/\sqrt{\hertz}}$ over a broad frequency bandwidth, independent of DC-offset currents. A sensitivity analysis reveals that the performance of our system is currently limited by intensity or polarization noise of the probe laser, but if mitigated, a current sensing sensitivity of about $\unit{4.3}{\pico\ampere/\sqrt{Hz}}$ should be achievable, while maintaining full galvanic isolation from the target circuit. 

Besides the investigation of fundamental research questions,  our findings will lay the foundation for novel approaches to integrating SERF magnetometers in feedback-controlled systems, enabling unprecedented precision and noise reduction in current control for next-generation technologies.

\section{Magnetometer dynamics}
\label{sec:mag_dynamics}

Optical pumped magnetometers work by detecting changes in light polarisation caused by the interaction of optically pumped alkali atoms with the magnetic field being measured. In the most sensitive SERF regime, relaxation processes are minimized through high atomic density and low magnetic field conditions. This regime is defined by the fact that the rate of spin-exchange collisions $\Gamma_{SE}$ is much faster than the Larmor frequency $\omega_L$ at which the atomic spins in the vapour cell precess in the magnetic field $\bm{B}$ 
\cite{li2018serf, Savukov2017}.
Under these conditions, the dynamics of the atomic spin polarization $\bm{P}$ can be described 
using a Bloch-type equation \cite{li2018serf, Savukov2017}, 
\begin{align}
    \frac{\text{d} \bm{P}}{\textbf{d}t}
    =
    \frac{1}{q} \left[
        \gamma_e \bm{P} \times \bm{B}
        +
        \Gamma_{\text{OP}} \left(\bm{\zeta} - \bm{P} \right)
        -
        \Gamma_{\text{rel}} \bm{P}
    \right]. \label{eq:Bloch}
\end{align}
Here, ${\bm{P} = \left< \bm{S} \right> / S}$
denotes the degree of spin polarization,
where $\left< \bm{S} \right>$ represents the ensemble-averaged spin
and $S$ the spin in the state targeted by optical pumping.
Moreover, $q$ is the nuclear slowing-down factor \cite{li2018serf, Savukov2017}, 
$\gamma_e$ represents the electron 
gyromagnetic ratio, $\Gamma_{\text{OP}}$ the optical pumping rate, and ${\Gamma_{\text{rel}}}$ is the total spin relaxation rate.
In an ideal SERF magnetometer, ${\Gamma_{\text{rel}}}$ is determined by the spin-destruction collision rate $\Gamma_{\text{sd}}$ \cite{li2018serf, Savukov2017}. The vector $\bm{\zeta}$ points in the direction of the pumping beam and its absolute value 
${0 \leq \zeta \leq 1}$ defines the degree of circular polarization of the pumping photons.

The optical pumping rate 
$\Gamma_{\text{OP}}$ at a pump photon current density in the pump beam
$j_{\text{p}}$, is given by \cite{appelt1999light}
\begin{align}
    \Gamma_{\text{OP}} = 
    j_{\text{p}} r_e c f 
    \frac{\Delta \nu /2}{ \left( \Delta \nu /2\right)^2 + \left(\nu - \nu_0 \right)^2} .
    \label{eq:gammaOP}
\end{align}
Here, ${\nu_0}$ represents the frequency of the target atomic transition,
accompanied by a linewidth ${\Delta \nu}$, which for SERF magnetometers is typically inhomogeneously (pressure) broadened to several GHz. 
Additionally, $r_e = \unit{2.8 \times 10^{-15}}{\metre}$ denotes the classical electron radius,
$c$ the speed of light, and $f$ the oscillator strength of the transition in use, which is often chosen to be an alkaline D1 transition. 

\begin{figure*}[tbp]
	\centering
	\includegraphics[width=17cm]{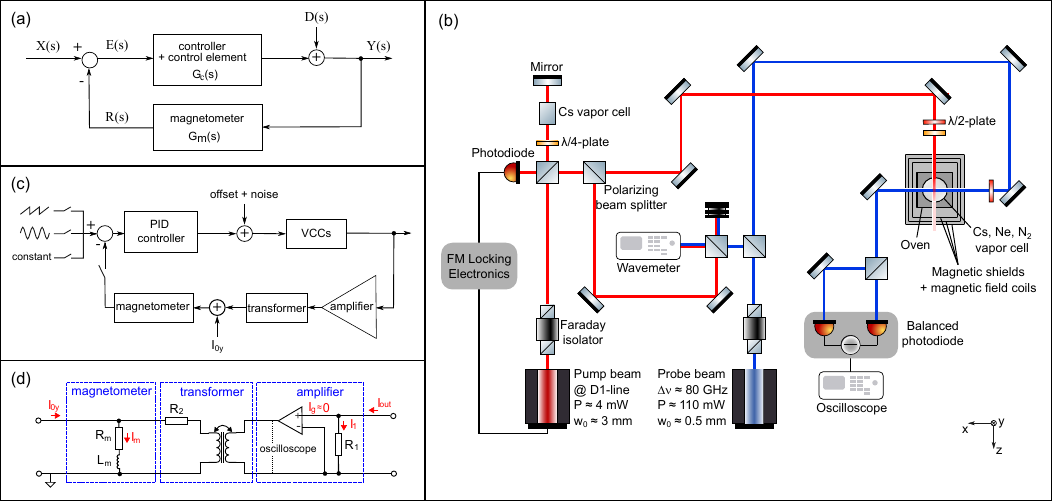}
	\caption{\textbf{Experimental setup.} 
	(a) Electronic scheme of a magnetometer placed in the feedback loop of a control circuit. $X(s)$, $Y(s)$, $R(s)$, $E(s)$ and $D(s)$ denote the input-,
    output-, feedback-, error-, and disturbance signal in the complex frequency domain. $G_c(s)$ and $G_m(s)$ denote the transfer function of the controller + control element and the magnetometer. 
    (b) The magnetometer setup is composed of a shielded vapor cell and a laser system for pump-probe detection. The pump laser is stabilized to the cesium D1 line near \unit{894.953}{\nano \meter} using frequency modulation spectroscopy. The detuning of the probe laser relative to the pump laser is determined with a wavelength meter. The vapor cell is enclosed within a four-layer magnetic shielding, which also houses a foil-based precision magnetic field coil system. A balanced photodiode, in combination with a polarizing beam splitter in the probe beam path after the vapor cell, detects changes in the polarization angle of the linearly polarized probe light, to produce the magnetometer signal. 
    (c) Signal flow diagram of the feedback loop for current noise suppression of a voltage-controlled current source (VCCS). A voltage offset and noise can be added to the PID output to set the VCCS operation current and noise characteristics. The noisy current is amplified and AC-filtered via an injection transformer. The remaining, amplified AC-current noise is transformed to a magnetic field noise at the magnetometer and subsequently (via the probe beam polarization) to a voltage noise at a photodiode. Finally, this feedback signal is subtracted from the input and fed as error signal to the proportional-integral-derivative (PID) controller. Between the injection transformer and the magnetometer, the DC-current $I_{y0}$, required for close-to-zero-field operation of the magnetometer, is superimposed to the current noise. The circuit can be operated in open- or closed loop, and the input being chosen as constant voltage, voltage ramp or sinusoidal oscillation, as used for measuring the noise suppression, and the DC- and AC-response of the magnetometer, respectively. 
    (d) Electronic circuit diagram of the feedback branch. The current from the power source feeds the working circuit via resistor $R_1$, which is used to tap the current and current noise as a voltage signal. This signal is then amplified and transferred to the secondary circuit of the transformer, before it is fed to the magnetometer coil (resistor $R_m$ and inductance $L_m$) via a resistor $R_2$. 
	}
	\label{fig:setup}
\end{figure*}

The fundamental concept of the magnetometer lies in the fact that a magnetic field $\bm{B}$ induces dynamics in ${\bm{P}}$, which is then measured to extract information about $\bm{B}$. 
For the following considerations, we assume the orthogonal magnetometer configuration (as realized in our experiment, cf.~Fig.~\ref{fig:setup}) in which the probe beam, the measured field and the pump beam are mutually orthogonal to each other, and denote their corresponding directions by $x$, $y$ and $z$, respectively. Then, $\bm{B}=B\bm{e}_y$ and $\bm{\zeta} = \zeta \bm{e}_z$, with $\bm{e}_i$, $i\in x,y,z$ denoting the unit vector in the corresponding direction. 
Operating the magnetometer for temporally slowly varying DC-like signals of $\bm{B}$, Eq.~\eqref{eq:Bloch} is well satisfied by the stationary solution 
${\mathrm{d}\bm{P} / \mathrm{d} t = 0}$.
Under the conditions 
${B_x = B_z = 0}$, ${B_{y} \neq  0}$ and $\zeta=1$  the stationary solution becomes
\begin{align}
    \begin{split}
    P_x &= - \frac{\gamma_e \Gamma_{\text{OP}} B_y}{\left(\Gamma_{\text{OP}} + \Gamma_{\text{rel}} \right)^2 + \gamma_{e}^2 B_{y}^2}\\
    P_y &= 0 \\
    P_z &= \frac{\Gamma_{\text{OP}} \left(\Gamma_{\text{OP}} + \Gamma_{\text{rel}} \right) }{\left(\Gamma_{\text{OP}} + \Gamma_{\text{rel}} \right)^2 + \gamma_{e}^2 B_{y}^2}.   
    \end{split}
    \label{eq:steady_state}
\end{align}

The component $P_x$ can then be measured using the Faraday effect, which induces a rotation of the polarization vector of a linearly polarized probe beam pointing along $\bm{e}_x$.
This rotation angle $\Delta \phi$ is given by \cite{ledbetter2008spin, happer1967effective}
\begin{align}
    \Delta \phi = - \frac{1}{2}
    r_e c f l n
    \frac{\left(\nu - \nu_0 \right)}{ \left( \Delta \nu /2\right)^2 + \left(\nu - \nu_0 \right)^2}\
    P_x.
    \label{eq:polarization_rotation}
\end{align}
Here, $n$ is the atomic density and $l$ the interaction length for the probe beam and the atoms within the cell. 

Again under the conditions ${B_x = B_z = 0}$, ${B_{y} \neq  0}$, and ${\zeta=1}$ and further assuming a small value for ${\int_{0}^t (\gamma_e / q) B_{y}(\tau)\ \mathrm{d} \tau}$, Appendix~\ref{ap:BlochAnaSmall} shows that the AC-response of the time dependent polarization $P_x (t)$ behaves as a first order low-pass filter.
The corresponding transfer function is
\begin{align}
    G_{m}(s) &= \frac{\tilde{P}_{x}(s)}{\tilde{B}_{y}(s)} = \frac{K_m}{1 + T_m\ s}\\
   |K_m| &= \frac{\gamma_e \Gamma_{\text{OP}}}{\left( \Gamma_{\text{OP}} + \Gamma_{\text{rel}} \right)^2}\\
    T_m &= \frac{q}{\Gamma_{\text{OP}} + \Gamma_{\text{rel}}},
    \label{eq:ACMag}
\end{align}
where ${\tilde{P}_x(s) = \mathcal{L} \left[ P_x(t) \right](s)}$ and ${\tilde{B}_y(s) = \mathcal{L} \left[ B_y(t) \right] (s)}$ are the Laplace transforms of ${P_x(t)}$ and ${B_y(t)}$, respectively. 
Here, ${s}$ denotes the complex valued frequency. The frequency response of the system is obtained by evaluating ${G_m(s)}$ at ${s = i \omega}$, with the real valued frequency $\omega$.

For the study presented here, we place the SERF magnetometer as a sensor element in the feedback branch of a closed-loop control system, as illustrated in Fig.~\ref{fig:setup}a. 
Here, $X(s)$, $Y(s)$, $R(s)$ and $E(s)$ denote the input-, output-, feedback- and error-signal in the s-domain, respectively. 
Additionally, a disturbance $D(s)$ is taken into account. 
In this scenario, as Appendix~\ref{ap:TransferFunc} demonstrates, the relationship between the input ${X(s)}$ and the disturbance ${D(s)}$ with the output ${Y(s)}$ is given by
\begin{align}
    \begin{split}
    Y(s) &= G_X (s) X(s) + G_D (s) D(s)\\
    G_X (s) &=  \frac{G_c(s)}{1 + G_m(s) G_c (s)}\\
    G_D (s) &=  \frac{1}{1 + G_m(s) G_c (s)}. 
    \label{eq:trans_CL}
    \end{split}
\end{align}
Here, ${G_c (s)}$ represents the transfer function of the controller and control element,
while ${G_X (s)}$ and ${G_D (s)}$ describe the system response to the input and the disturbance, respectively.
As an example, ${G_c (s)}$ may take the form of an ideal PID controller combined with a (parasitic) time constant of the control element, expressed as
\cite{keviczky2019control}
\begin{align}
    G_c(s) = K \left(1 + \frac{1}{T_i\ s} + T_d\ s \right) \frac{1}{1 +T_e\ s}.
\end{align}
Here, $K$ represents the overall gain of the controller in combination with the control element. 
The parameter $T_i$ denotes the time constant of the integrator within the controller, while $T_d$ corresponds to the time constant of the differentiator. 
Finally, $T_e$ is the time constant associated to the control element.

Beyond the DC- and AC-characteristics of the magnetometer, a key performance metric is represented by its sensitivity. 
A practical measure of sensitivity is the amplitude spectral density (ASD) of the device noise floor \cite{bhattacharyya2018handbook}, 
as a signal must exceed this threshold to be detectable.
In a classical measurement scheme, a limit for this noise floor is given by the standard quantum limit (SQL) of the 
spin polarization measurement. This is determined by the spin projection noise, 
arising from the non-commutativity of two spin components ${S_j}$ and ${S_k}$, where ${j \neq k}$ \cite{savukov2005tunable}.
In the context of an optically pumped magnetometer, spin projection noise leads to a root-mean-square deviation $\Delta B_{\text{rms}}$ of the magnetic field amplitude \cite{savukov2005tunable}. 
When normalized by the square root of the measurement bandwidth ${\Delta f}$, this uncertainty defines the amplitude spectral density 
characterizing the magnetometer sensitivity at the SQL, given by \cite{savukov2005tunable}
\begin{align}
    \frac{\Delta B_{\text{rms}}}{\sqrt{\Delta f}} = \frac{q}{\gamma_e} \sqrt{\frac{\Gamma_{\text{rel}}}{N}}.
    \label{eq:SQL}
\end{align}
Here, the measurement bandwidth is ${\Delta f = 1/t_{\text{meas}}}$, where ${t_{\text{meas}}}$ is the measurement time, and $N$ represents the number of atoms in the probe volume of the spin-polarized ensemble.

\section{Experimental methods}
\label{sec:exp_setup}

\subsection{Magnetometer setup}
\label{subsec:mag_setup}

Figure.~\ref{fig:setup}b sketches the experimental setup of the magnetometer. It is composed of a shielded Cs vapor cell and a laser-based pump-probe system for spin polarization and detection. 
Both the pump and the probe laser are external-cavity diode lasers in Littrow configuration \cite{ricci1995compact}.
The pump laser is stabilized to the $\ket{6S_{1/2},F=3} \rightarrow \ket{6P_{1/2},F'=4}$ transition of the cesium D1-line using frequency-modulation spectroscopy \cite{bjorklund1983frequency}.
The probe laser is blue-detuned with respect to the pump laser by about ${\unit{80}{\giga \hertz}}$, as confirmed with a  wavelength meter. Unless stated otherwise, the optical power of the pump and probe beams are $\unit{4}{\milli \watt}$ and $\unit{110}{\micro \watt}$, respectively. 
The corresponding beam waists ($1/e^2$-intensity radius) are $\unit{3}{\milli \metre}$ for the pump beam and $\unit{0.5}{\milli \metre}$ for the probe beam.

The cesium vapor cell used for magnetometry is placed within a four-layer magnetic shielding (Twinleaf MS-2).
The spherical cell of $\unit{25.4}{\milli \metre}$ diameter is filled with a mixture of $\unit{600}{Torr}$ of neon, serving as buffer gas, and $\unit{50}{Torr}$ of nitrogen, serving as quenching gas. 
During experiments, the cell is heated to ${T = \unit{383}{\kelvin}}$, defining the vapor pressure and thus the atom density of the cesium that is encapsulated into the cell
\cite{alcock1984vapour}.
A precision magnetic field coil system is integrated within the shielding. 
This system enables \textit{inter alia} the definition of the magnetic flux density components ${B_{x,y,z}}$ 
through three separate currents ${I_{x,y,z}}$, applied to respective coils, according to
\begin{align}
    \begin{split}
        B_{x,y} &= \beta_{x,y}\ I_{x,y}\ \ \ \text{with}\ \ \ \beta_{x,y} = \unit{57.6}{nT/mA}\\
        B_{z} &=  \beta_{z}\ I_{z}\ \ \ \ \ \ \ \text{with}\ \ \ \beta_{z} = \unit{105}{nT/mA}.
        \label{eq:B_I_Dep}
    \end{split}
\end{align}
Prior to experiments, the residual magnetic field at the vapor cell position is nullified following the procedure outlined in Ref.~\onlinecite{zhao2019non}, resulting in the currents ${I_{k0}\ (k=x,y,z)}$, which are applied to the corresponding coils using an ultra-low noise current source from DM Technologies. SERF operation is only possible in these close-to-zero-field conditions.

\subsection{Electronic setup}
\label{subsec:electronic_setup}

\begin{table}[tbp]
\begin{center}
\caption{Current sources used in this study}
\begin{tabular}{ c  c  c } 
 current source & type & max. current\\
 \hline
 \hline
VCCS1 & HighFinesse Milliamp Line  & $\unit{20}{\milli \ampere}$ \\ 
VCCS2 & HighFinesse Compact Line  & $\unit{3}{\ampere}$ 
 \label{tab:CS}
\end{tabular}
\end{center}
\end{table}

While Fig.~\ref{fig:setup}a illustrated the general system design with a magnetometer placed in the feedback branch of a closed-loop control circuit, Fig.~\ref{fig:setup}c presents a more detailed system design specific to this study. Here, the controller is a commercial PID controller (SIM960 from Stanford Research Systems) and the control element is a commercial voltage-controlled current source (VCCS), which is meant to be stabilized with respect to current noise.
The setting of the DC-current offset is performed using the SIM960 internal adder, while an external adder allows for the injection of well-defined noise into the VCCS originating from a waveform generator (Agilent 33520B).
Within the scope of this study, we use two different current sources, as summarized in Tab.~\ref{tab:CS}.
Additionally, we use an injection transformer (B-LFT 100 Low-Frequency from Omicron Lab) combined with an upstream amplifier (DLPVA-100-BD from FEMTO Messtechnik GmbH) in the feedback-branch of the circuit, such that only the AC-signal (noise) from the VCCS is coupled to the magnetometer. In between, the DC-current $I_{0y}$ is superimposed, which is required to nullify the magnetometer DC-field. 

To distinguish the signal flow diagram in Fig.~\ref{fig:setup}c from the circuit diagram, we sketch the electronic design of the feedback branch in Fig.~\ref{fig:setup}d. 
The output current $I_{\text{out}}$ of the VCCS is converted to a voltage over $U = R_1 I_{\text{out}}$ and amplified, with ${R_{1} = \unit{10}{\ohm}}$.
The amplification factor is 20 or $\unit{40}{dB}$, depending on the specific measurement.
The output voltage of the amplifier is coupled to the injection transformer, which in turn connects to the resistance $R_2$ yielding a current to the magnetometer $B_y$-coil (illustrated here via a resistance $R_m$ and an inductivity $L_m$). 
The voltage transfer function of the transformer exhibits the characteristics of a band-pass filter, 
with a lower ${\unit{3}{dB}}$ cut-off frequency of about ${f_\text{L} = \unit{30}{\milli\hertz}}$, only slightly varying with $R_2$.  
The upper ${\unit{3}{dB}}$ cut-off frequency depends on $R_2$ and varies within the range 
${\unit{2.4}{\kilo\hertz} \leq f_H \leq \unit{800}{\kilo\hertz}}$ for $\unit{10}{\ohm} \leq R_2 \leq \unit{10}{\kilo\ohm}$ (for the measurements we use $R_2=0$), as determined by characterization measurements. 
The output is then coupled to the magnetometer. 
In addition, the zero current $I_{0y}$ is applied to cancel the $B_y$ field when no current 
from the voltage-controlled current source is applied to the magnetometer. 

To characterize the noise generated by the VCCS under consideration, the voltage drop across the resistor $R_1$ is recorded after the amplifier with a digital oscilloscope (PicoScope 5242D), 
capturing multiple time traces, each with a duration of ${\unit{1}{\second}}$ and a sampling rate of ${\unit{250}{\kilo \hertz}}$. 
The recorded voltage signals are converted to a current measurement using the specified gain of the amplifier and the value of $R_1$. 
For each of these current signals, we compute the double-sided amplitude spectral density ASD and the resulting spectra are finally averaged.

\section{Results}

\subsection{Magnetometer and setup characterization}
\label{subsec:results_characterization}

\begin{figure}[tbp]
	\centering
	\includegraphics[width=8.5cm]{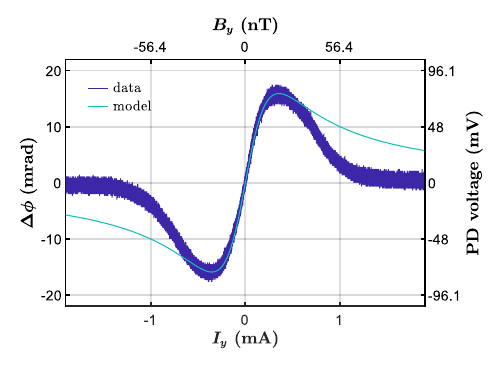}
	\caption{\textbf{DC-Magnetometer response.} 
	    Variation in the polarization angle (or, correspondingly, the voltage on the balanced photodiode) as measured as a function of the current $I_y$, which corresponds to the magnetic field ${B_y = \beta_y I_y}$. Measurement data (blue) is shown together with the model function from Eq.~\eqref{eq:steady_state} (turquoise). 
	}
	\label{fig:DCsignal}
\end{figure}

\subsubsection{Open-loop DC-response}
Figure~\ref{fig:DCsignal} shows the response of the magnetometer to a quasi-static, DC-like modulation of 
${B_y = \beta_y I_y}$, which is induced by $I_y$ using current source 1. 
To measure the DC-behavior of the magnetometer, the PID controller in Fig.~\ref{fig:setup}c 
is effectively bypassed by setting its transfer function to ${G_{\text{pid}}(s) = 1}$.
A sawtooth voltage signal is applied to the VCCS via the Agilent waveform generator, resulting in a modulation of $B_y$ with an amplitude ${A=\unit{112.8}{\nano \tesla}}$ and a frequency ${f=\unit{100}{\milli \hertz}}$, which is well above the transformers lower cut-off frequency. 
The principal form of the DC-response follows Eq.~\eqref{eq:polarization_rotation}, with $P_x$ given from the steady state solution in Eq.~\eqref{eq:steady_state}, and it clearly shows the linear response of the measured polarization rotation (PD voltage) with respect to the magnetic field $B_y$ (current $I_y$) close to zero-field conditions. 
The extrema of the DC-response occur at
${B_y = \pm (\Gamma_{\text{OP}} + \Gamma_{\text{rel}})/\gamma_e}$. From the extrema observed in Fig.~\ref{fig:DCsignal}, 
located at ${B_y \approx \pm \unit{20}{nT}}$, 
one can deduce that ${\Gamma_{\text{OP}} + \Gamma_{\text{rel}} \approx \unit{3.5 \times 10^3}{\second^{-1}}}$. However, the model function fails to describe the wings of the measured DC-response beyond the extreme position. This may be due to the breakdown of the SERF regime requirements or shortcomings in the Bloch model, which will be addressed in future studies. 

\begin{figure*}[tbp]
	\centering
	\includegraphics[width=17.0cm]{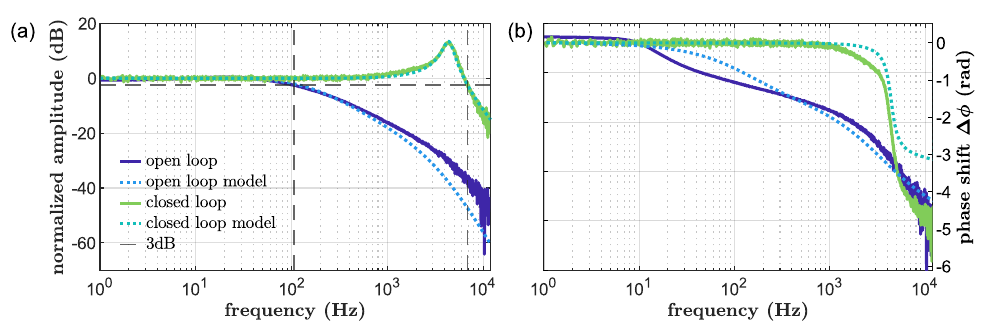}
	\caption{\textbf{AC-Magnetometer response.} 
        Characterization of the magnetometer transfer function in terms of Bode plots for (a) the amplitude response and (b) the phase response, as measured in open- and closed-loop configurations (solid lines). Theoretical predictions from Eq.~\eqref{eq:AC_response_openloop} and \eqref{eq:modelCL} are shown as colored dashed lines, based on fitting the open- and closed-loop amplitude responses simultaneously with equal weight. The closed-loop PID-controller parameters are determined using the Ziegler–Nichols method \cite{ziegler1942optimum}. The ${\unit{3}{dB}}$ cut-off frequency amounts to
        ${\unit{105}{\hertz}}$ in open-loop configuration and is extended to ${\unit{6.9}{\kilo\hertz}}$ in closed-loop configuration.
        }
	\label{fig:OL_CL}
\end{figure*}

\subsubsection{Open- \& Closed-loop AC-response:}
Figure~\ref{fig:OL_CL}a and \ref{fig:OL_CL}b characterize the AC-behavior of the system, depicting the amplitude and phase response of the magnetometer, which relates the signal ${R(s)}$ to the input ${X(s)}$ in Fig~\ref{fig:setup}b.
The blue curves in Fig.~\ref{fig:OL_CL}a and \ref{fig:OL_CL}b correspond to the open-loop configuration, where the PID controller is effectively bypassed and the feedback loop remains open. In this scenario, the system transfer function from ${X(s)}$ to ${R(s)}$ is determined by the cascaded responses of the VCCS, amplifier, transformer, and magnetometer, namely 
\begin{align}
\begin{split}
    G_{\text{vccs}}(s) = \frac{K_{\text{vccs}}}{1 + T_{\text{vccs}} s}, \quad
    G_a(s) = \frac{K_a}{1 + T_a s}, \\
    G_t(s) = \frac{2 D_t \omega_t s}{\omega_t^2 + 2 D_t \omega_t s + s^2},\quad 
    G_m(s) = \frac{K_m}{1 + T_m s}.
\end{split}
\end{align}
The overall system response is thus given by
\begin{equation}
    R(s) = G_{R,\text{open}}(s)\ X(s),
\end{equation}
where
\begin{equation}
    G_{R,\text{open}} (s) = G_{\text{vccs}}(s)\ G_{a}(s)\ G_{t}(s)\  G_{m}(s).
    \label{eq:AC_response_openloop}
\end{equation}

The green curves in Fig.~\ref{fig:OL_CL}a and \ref{fig:OL_CL}b represent the system
response when the feedback loop is closed and the PID controller is included.
Similar to the derivation in Appendix~\ref{ap:TransferFunc}, the system transfer function from X(s) to R(s) in Fig.~\ref{fig:setup}b is then modelled by
\begin{equation}
    G_{R,\text{closed}} (s)  = \frac{G_{\text{pid}}(s)\ G_{R,\text{open}}(s)}{1 + G_{\text{pid}}(s)\ G_{R,\text{open}}(s)}
    \label{eq:modelCL}
\end{equation}
with
\begin{equation}
    G_{\text{pid}} (s) = K_{\text{pid}} \left(1 + \frac{1}{T_i\ s} + T_d\ s \right).
\end{equation}

The amplitude and phase responses, shown in Fig.~\ref{fig:OL_CL}a and \ref{fig:OL_CL}b, correspond to ${|G_R(i\omega)|}$ and ${\arg(G_R(i\omega))}$ both in open- and closed loop. 
Besides the measurement results, these figures include theoretical predictions as obtained from fits of Eqs.~\eqref{eq:AC_response_openloop} and \eqref{eq:modelCL} to the open- and closed-loop amplitude responses simultaneously with equal weight. The following parameters are extracted: the bandpass characteristics of the injection transformer, with ${D_t = 110.9}$ and ${\omega_t = \unit{27.8}{rad/s}}$, yielding an upper cut-off frequency of ${f_H = \unit{1.01}{\kilo \hertz}}$. The magnetometer time constant is fitted to ${T_m = \unit{1.1}{ms}}$, resulting in a ${\unit{3}{dB}}$ cut-off frequency of ${1/(2\pi\times T_m)=\unit{144}{\hertz}}$, which aligns closely to the experimental measured value of \unit{105}{\hertz} (cf.~Fig.~\ref{fig:OL_CL}a).
The time constants ${T_{\text{vccs}} = 1/(2 \pi \times \unit{10.5}{\kilo \hertz})}$ and ${T_a = 1/(2 \pi \times \unit{100}{\kilo \hertz})}$ are \textit{not} fitted, but stem 
from measurements and from the amplifier datasheet, respectively.
The overall system gain is normalized to ${K = K_{\text{vccs}} K_a K_m = 1}$. The PID time constants $T_i = {\unit{111}{\micro \second}}$ and $T_d = {\unit{28}{\micro \second}}$, 
as well as its gain $K_{\text{pid}} = 162$ match the experimental values, as determined using the Ziegler–Nichols tuning method \cite{ziegler1942optimum}. 

Besides their general behaviour, the data in Fig.~\ref{fig:OL_CL}a and \ref{fig:OL_CL}b indicate an enhancement in bandwidth, 
as observed from the ${\unit{3}{dB}}$ cut-off frequency. 
Specifically, the cut-off frequency increases from ${\unit{105}{\hertz}}$ in the open-loop case, where the PID controller is bypassed, 
to ${\unit{6.9}{\kilo\hertz}}$ in the closed-loop case, where the PID controller is incorporated.

\begin{figure}[tbp]
	\centering
	\includegraphics[width=8.5cm]{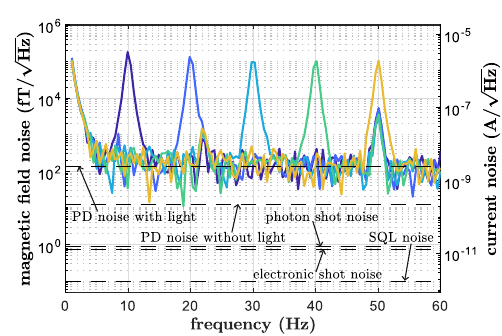}
	\caption{\textbf{Magnetometer sensitivity.} Amplitude spectral density as measured with a lock-in amplifier at the output of the magnetometer for different modulation frequencies (different colors) of $I_y$. The measurement is performed in open loop with the output of the VCCS directly connected to the y-coil of the magnetometer. The measurements determine the sensitivity-defining noise floor to be ${\unit{2.46}{\nano\ampere/\sqrt{\hertz}}}$, corresponding to ${\unit{139}{\femto\tesla/\sqrt{\hertz}}}$ (PD noise).
	    Analyzing various technical and physical noise sources (black dashed lines) suggests that the magnetometer is limited by the illuminated photodiode. 
	    The fundamental noise floor of our setup is given by the standard quantum limit (SQL), which yields ${\unit{1.75}{\pico\ampere/\sqrt{\hertz}}}$, 
	    corresponding to ${\unit{0.1}{\femto\tesla/\sqrt{\hertz}}}$.
        }
	\label{fig:sensitivity}
\end{figure}

\subsubsection{Sensitivity:}
The sensitivity achieved by our magnetometer is illustrated in Fig.~\ref{fig:sensitivity}. 
For characterization, $B_y$ is modulated at various frequencies with an amplitude ${A = \unit{80}{\pico \tesla}}$. 
The PID controller in Fig.~\ref{fig:setup}c is bypassed, and the measurement is performed in open loop.
The amplitude spectral density (ASD) of the magnetometer output signal (output voltage of the balanced photodiode in Fig.~\ref{fig:setup}a) 
is obtained using a lock-in amplifier (HF2LI from Zürich Instruments). 
This measures the signal power within a frequency interval 
${[f_{r} - \Delta f_{NEP},  f_{r} + \Delta f_{NEP}]}$, 
where the noise equivalent power bandwidth is ${\Delta f_{NEP} = \unit{0.28}{\hertz}}$ in our measurements. 
The reference frequency $f_r$ is tuned over the frequency range of interest using 120 sample points. 
The signal power within ${[f_{r} - \Delta f_{NEP},  f_{r} + \Delta f_{NEP}]}$ 
is characterized by mixing the output signal with a reference signal of frequency $f_r$ 
and extracting its DC-component using a fourth-order low-pass filter with a cut-off frequency of ${\Delta f_{NEP}}$. 
From the measured power and ${\Delta f_{NEP}}$, the power spectral density (PSD) and ASD are determined \cite{bhattacharyya2018handbook}. 
The output voltage ASD is referred to the input magnetic field ASD by the relation of the output voltage amplitude to input
magnetic field amplitude of ${\unit{80}{\pico \tesla}}$. 
Since the magnetic field ${B_y = \beta_y I_y}$ is directly given by the current $I_y$, 
the current ASD can also be determined accordingly.

Figure~\ref{fig:sensitivity} presents the amplitude spectral density (ASD) of both $B_y$ and $I_y$. 
The noise floor, which serves as a measure of the magnetometer sensitivity, is found to be ${\unit{139}{\femto\tesla/\sqrt{Hz}}}$ or ${\unit{2.46}{\nano\ampere/\sqrt{Hz}}}$.
Additionally, Fig.~\ref{fig:sensitivity} shows the noise levels corresponding to the illuminated- and dark balanced photodiode, as well as the noise contributions expected from the photon shot noise of the probe beam and
the electronic shot noise of the photo current \cite{bachor2019guide}. 
The noise level of the illuminated photodiode closely matches the noise floor observed in our measurements, 
indicating that the magnetometer sensitivity is primarily limited by this factor. 
Since the noise level of the illuminated photodiode is higher than that of the dark photodiode, this suggests that the dominant noise source originates from the probe beam, which could be attributed to either intensity or polarization fluctuations.

\begin{figure*}[!t]
	\centering
	\includegraphics[width=17.0cm]{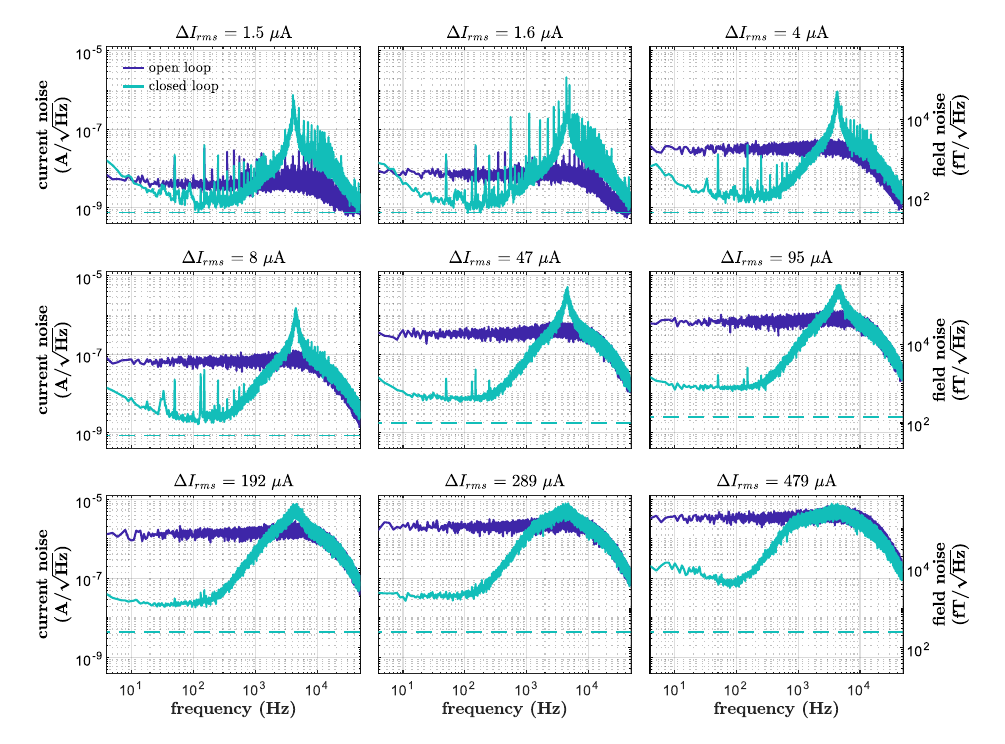}
	\caption{\textbf{Noise suppression for different noise levels.} 
        Current noise spectra in open loop (blue) and closed loop (turquoise) as measured for different levels of applied (open-loop) noise $\Delta I_{\text{rms}}$ to VCCS1 from Tab.~\ref{tab:CS}. The horizontal dashed line in each plot corresponds to the noise level of the voltage amplifier in the configuration used for the closed-loop measurements, that is with gain set to 40$\,$dB and the appropriate voltage scale at the oscilloscope, as measured for a short-circuited input signal and converted into current noise (cf.~Appendix~\ref{ap:Devices_noise}). All spectra lie above the lines, meaning that we are not limited by the noise of the measurement system itself.
        }
	\label{fig:an_dep_irms}
\end{figure*}

The optimal noise floor is determined by the SQL as expressed in Eq.~(\ref{eq:SQL}), yielding 
${\Delta B_{\text{rms}} / \sqrt{\Delta f} = \unit{0.1}{fT/\sqrt{Hz}}}$. 
In this context, the best slow-down factor is $q=8$~\cite{ledbetter2008spin, appelt1999light}. 
Furthermore, in an ideal Cs SERF magnetometer, the relaxation rate is determined by Cs--Cs spin-destruction collisions and is given by 
${\Gamma_{\text{rel}} = n \bar{v}_{\text{th}} \sigma_{\text{sd}}}$  \cite{allred2002high}. 
Here the spin-destruction cross section is 
${\sigma_{\text{sd}} = \unit{2 \times 10^{-16}}{cm^2}}$ \cite{allred2002high, bhaskar1980spin}, 
$\bar{v}_{\text{th}}$ denotes the average relative thermal velocity and $n$ is the atomic number density, both evaluated at the experimental atomic temperature $T=\unit{383}{K}$.  
Moreover, in Eq.~(\ref{eq:SQL}), $N$ is taken to be $N = nV$, 
assuming an attainable probe volume of ${V = \unit{1}{cm^3}}$.

\subsection{Current noise suppression}
\label{sec:current_noise_suppression}

Now we turn towards measurements regarding the stabilization of a VCCS using the magnetometer. Here, we investigate the dependency of the current noise suppression on different experimental parameters, such as the applied (open-loop) noise level and the DC-offset of the stabilized current.

Figure~\ref{fig:an_dep_irms} presents the measured current noise in terms of its ASD in both open- and closed-loop configurations for different noise levels. 
The noise is artificially generated by superimposing white noise onto the current of VCCS1 from Tab.~\ref{tab:CS}. The magnitude of this artificial noise is characterized by the root-mean-square value of the open-loop current $\Delta I_{\text{rms}}$, as determined from the measured time-resolved voltage drop at $R_1$.

\begin{figure}[tbp]
	\centering
	\includegraphics[width=8.5cm]{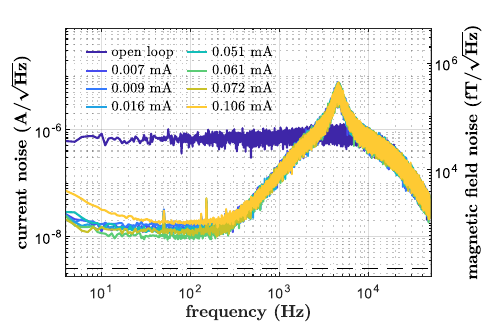}
	\caption{\textbf{Noise suppression for different DC-currents.}
        Current noise spectra for the $20\,$mA source (VCCS1) at different stabilization currents, corresponding to different voltage offsets from the PID controller. For all measurements, we use $\Delta I_{\text{rms}}=47\,\mu$A and a $40\,$dB gain at the voltage amplifier. Similar to Fig.~\ref{fig:an_dep_irms}, the horizontal dashed line corresponds to the noise level of the amplifier and oscilloscope used for measuring the current noise. All spectra lie above the line, meaning that we are not limited by the noise measurement system itself.
	}
	\label{fig:an_dep_offset}
\end{figure}

\begin{figure}[tbp]
	\centering
	\includegraphics[width=8.5cm]{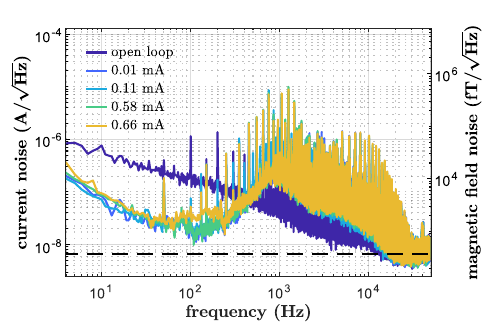}
	\caption{\textbf{Noise-suppression of commercial source.} 
        Current noise spectra for the $3\,$A source (VCCS2) at different stabilization currents, corresponding to different voltage offsets from the PID controller. For these measurements, no artificial noise is applied and the gain at the voltage amplifier is set to $20\,$dB. Similar to Fig.~\ref{fig:an_dep_irms} and \ref{fig:an_dep_offset}, the horizontal dashed line corresponds to the noise level of the measurement system (amplifier and oscilloscope). All spectra lie above the line, meaning that we are not limited by the noise of the measurement devices.
	}
	\label{fig:3A_dep_offset}
\end{figure}

As shown in Fig.~\ref{fig:an_dep_irms}, increasing $\Delta I_{\text{rms}}$ enhances current noise suppression up to almost two orders of magnitude. 
Furthermore, this analysis provides a measure for the lowest noise level that can be effectively suppressed. 
At the lowest noise $\Delta I_{\text{rms}}=\unit{1.5}{\micro \ampere}$, the noise suppression in closed loop is only slightly below the open-loop level, 
indicating the threshold of detectable noise. 
This corresponds to an open-loop noise floor of ${\unit{4.5 \times 10^{-9}}{A/\sqrt{Hz}}}$, or equivalently to ${\unit{250}{fT/\sqrt{Hz}}}$,
which is close to the characterized sensitivity of ${\unit{140}{fT/\sqrt{Hz}}}$ shown in Fig.~\ref{fig:sensitivity}. In addition, Fig.~\ref{fig:an_dep_irms} (and also Fig.~\ref{fig:an_dep_offset} and \ref{fig:3A_dep_offset}) show horizontal lines indicating the noise threshold of the measurement system itself. This depends on the gain settings on the amplifier and the scale setting at the oscilloscope and has been characterized for a short-circuited input signal ($R_1=0$) to the amplifier and converted into current noise (cf.~Appendix~\ref{ap:Devices_noise}). 

To demonstrate the noise suppression at different DC-offset currents and thus the functionality of the AC-coupling induced by the injection transformer, as shown in Fig.~\ref{fig:setup}c and Fig.~\ref{fig:setup}d,
we measured the current noise from source 1 at different DC-current levels. 
Figure~\ref{fig:an_dep_offset} presents the current noise suppression for various DC-currents at a noise level of ${\Delta I_{\text{rms}} = \unit{47}{\micro \ampere}}$, which is applied as described above.
The results show that the attenuated current noise in closed-loop configuration remains largely independent of the stabilization current DC-component, 
as long as it stays below a certain threshold of ${\unit{72}{\micro \ampere}}$. 
Beyond this threshold, the noise starts to increase in the low-frequency regime. 
This behavior can be attributed to the non-ideal characteristics of the transformer, which does not completely eliminate low-frequency current components but only suppresses them.

The same measurement was repeated with VCCS2 from Tab.~\ref{tab:CS}, this time without applying any additional noise. In fact, characterization measurements (cf.~Appendix~\ref{ap:Devices_noise}) show that the intrinsic noise level of this source reaches $\approx 2 \times 10^{-7}\,$A/$\sqrt{\text{Hz}}$, which is higher than the sensitivity of our setup and can therefore be regulated. The measured spectra are displayed in Fig.~\ref{fig:3A_dep_offset}, which again confirms that the current noise reduction is independent of the applied offset. However, this time the regulation only involves the noise of the current source itself, going beyond a mere proof of principles, but demonstrating the realization of a system with proper technological applications.

\section{Discussion and Conclusion}

Using a SERF magnetometer within the feedback branch of a closed-loop feedback system designed to suppress the noise of a voltage-controlled current source offers the advantage of perfect galvanic isolation as compared to standard transimpedance amplifiers.

Furthermore, the circuit we propose for AC-coupling, featuring an ultra-low lower cutoff frequency, effectively enables the definition of the DC-current supplied by the current source. We validated its functionality with two different current sources, each operating within totally different DC-current ranges.

In the low-frequency regime, our setup provides significant current noise suppression. However, the achieved sensitivity of
$\unit{2.46}{\nano\ampere/\sqrt{\hertz}}$, corresponding to
$\unit{139}{\femto\tesla/\sqrt{\hertz}}$, is higher than that of ultra-low-noise transimpedance amplifiers \cite{djekic2021440}.
Nevertheless, as demonstrated in this study, our setup was primarily limited by technical noise, particularly polarization and intensity noise of the probe beam. Eliminating these technical noise sources would enable operation at the standard quantum limit, yielding a magnetometer sensitivity of $\unit{0.24}{\femto\tesla/\sqrt{\hertz}}$, which corresponds to a current measurement sensitivity of $\unit{4.25}{\pico\ampere/\sqrt{\hertz}}$. This level of sensitivity is comparable to the Johnson noise of the coil system resistance, given by ${\Delta I_{\text{rms}} / \sqrt{\Delta f} = \unit{55}{pA/\sqrt{Hz}}}$ at room temperature.
Moreover, further improvements in sensitivity are, in principle, attainable. By increasing the coupling parameter $\beta_y$ and employing higher-resistance components to suppress Johnson noise contributions, it should be possible to achieve sensitivities in the ${\unit{}{fA/\sqrt{Hz}}}$ regime for current noise measurement and suppression.

\section*{Acknowledgements}
This work was supported by the German Federal Ministry of Economic Affairs and Energy through the WIPANO-program under project. no. 03THWBW001 and by the Deutsche Forschungsgemeinschaft through FOR 5413. 

\appendix

\section{Dynamics for $B_x = B_z = 0$}
\label{ap:BlochAnaSmall}

For ${B_x = B_z = 0}$ and a perfectly circular polarized pump beam aligned along the z-axis,
Eq.~\eqref{eq:Bloch} couples $P_x$ and $P_z$ as follows
\begin{align}
    \begin{split}
        \dot{P}_{x} &= - \Omega(t) P_z - \eta P_x \\
        \dot{P}_{z} &=  \Omega(t) P_x - \eta P_z + \xi ,
    \end{split} 
\end{align}
with ${\Omega (t) = \gamma_e B_{y} (t) / q}$, ${\xi = \Gamma_{\text{OP}}/q }$,
and ${\eta = (\Gamma_{\text{OP}} + \Gamma_{\text{rel}})/q}$.
For the homogeneous differential equation system a fundamental matrix $\Psi (t)$ can be found:
\begin{align}
    \Psi (t) =
    e^{-\eta t}
    \left( 
        \begin{matrix}
            \cos \left( \int_{0}^{t} \Omega (\tau)\ \mathrm{d} \tau \right) &  - \sin \left( \int_{0}^{t} \Omega (\tau)\ \mathrm{d} \tau \right)\\
            \sin \left( \int_{0}^{t} \Omega (\tau)\ \mathrm{d} \tau \right) & \cos \left( \int_{0}^{t} \Omega (\tau)\ \mathrm{d} \tau \right)
        \end{matrix}
    \right).
    \label{eq:fundSysGen}\nonumber
\end{align}
Assuming ${\int_{0}^{t} \Omega (\tau)\ \mathrm{d} \tau}$ to be small, this matrix can be linearized:
\begin{align}
     \Psi (t) = 
     e^{-\eta t}
      \left( 
        \begin{matrix}
            1 & - \int_{0}^{t} \Omega (\tau)\ \mathrm{d} \tau \\ 
            \int_{0}^{t} \Omega (\tau)\ \mathrm{d} \tau & 1
         \end{matrix}
      \right).
\end{align}
Under the  already made assumption of small ${\int_{0}^{t} \Omega (\tau)\ \mathrm{d} \tau}$,
variation of constants 
leads to the general solution of the non-homogeneous differential equation
\begin{align}
    \begin{split}
    \left( 
        \begin{matrix}
            P_x \\ 
            P_z
         \end{matrix}
      \right)
      &=
     \Psi (t) 
     \cdot
     \left( 
        \begin{matrix}
            c_{x} \\ 
            c_{z}
         \end{matrix}
      \right) \\
      &+
      \Psi (t) \cdot
      \bigintsss_{0}^{t}
      \xi\ e^{\eta \tau_1 }
       \left( 
        \begin{matrix}
            \int_{0}^{\tau_1} \Omega (\tau_2 )\ \mathrm{d} \tau_2 \\ 
            1
         \end{matrix}
      \right)\
      \mathrm{d}\tau_1.
      \end{split}.
\end{align}
Here, the constants $c_{x,z}$ are defined by the initial conditions.
According to Eq.~\eqref{eq:polarization_rotation} the measured output signal of the magnetometer is proportional to $P_x$.
Assuming $\Omega(t) = 0$ for all $t < 0$ and
the equilibrium conditions 
${P_x(t=0) = 0}$ and ${P_z(t=0) = \xi / \eta}$, it holds
\begin{align}
    P_x = -\frac{\xi}{\eta} \int_{0}^{t} \Omega(\tau) e^{-\eta (t - \tau)}\ \mathrm{d} \tau,
    \label{eq:solPxAna}
\end{align}
which can be understood as a convolution of $\Omega (t)$ with ${e^{-\eta t}}$.
Thus, by introducing the Laplace transforms 
\begin{align}
    \Tilde{P}_x (s) = \mathcal{L} \left[ P_x (t) \right] (s)\ \ \text{and}\ \ 
    \Tilde{\Omega}_x (s) = \mathcal{L} \left[ \Omega (t) \right] (s),
\end{align}
the frequency domain expression of Eq.~\eqref{eq:solPxAna} becomes
\begin{align}
    \Tilde{P}_x (s) = \frac{-\xi / \eta}{s + \eta}\ \Tilde{\Omega}_x (s).
    \label{eq:solPxAnaFreqDomain}
\end{align}
Equation~\eqref{eq:solPxAnaFreqDomain}
describes the simple behaviour of a first order low-pass filter with $\eta$
being the cutoff angular frequency.

\section{Working with transfer functions}
\label{ap:TransferFunc}
For a linear and time-invariant system, there are two equivalent ways of describing the system output as a function of an arbitrary input.  
On the one hand, the system can be described in the time domain, where it is characterized by the impulse response $ g(t)$.  
In this case, the system output $y(t)$ is given by the convolution of the input $x(t)$ and $g(t)$, namely  
$y(t) = (g*x)(t)$.
On the other hand, the system can be described in the (complex-valued) frequency domain, where it is characterized by the transfer function $G(s)$,  
with $s = \sigma + i \omega$ being the complex frequency and $G(s) = \mathcal{L}\left[g(t)\right](s)$ the Laplace transform of $g(t)$.  
In the frequency domain, the relationship between input and output is given by  $Y(s) = G(s) X(s)$, where  $Y(s)$ and $X(s)$ are the Laplace transforms of $y(t)$ and $x(t)$, 
respectively.

If two subsystems with $G_1 (s)$ and $G_2(s)$ are connected in series the total transfer function is given by ${G(s) = G_1(s) G_2(s)}$.
Moreover, if two signals are added in the time domain, they are also added in the frequency domain.
Thus, for Fig.~\ref{fig:setup}a, the following holds:
\begin{align}
Y(s) &= G_c(s) E(s) + D(s), \\
E(s) &= X(s) - R(s) = X(s) - G_m(s) Y(s),
\end{align}
and thus,
\begin{align}
Y(s) = \frac{G_c(s)\ X(s) + D(s)}{1 + G_m(s)\ G_c(s)},
\end{align}
which defines the two transfer functions in Eq.~\eqref{eq:trans_CL}.

\section{Devices noise characterization}
\label{ap:Devices_noise}

We present the characterisation of the noise produced by all the devices used in the feedback loop designed to stabilise the current.
These measurements are of particular interest because they enable us to determine whether the reduction in current noise is limited by these devices. Specifically, all noise measurements presented in Sec.~\ref{sec:current_noise_suppression} are performed by detecting time-traces of the signal of interest with an oscilloscope (PicoScope 5242D), after amplification with a voltage amplifier (Femto DLPVA-100-BD). Therefore, these devices must be characterised to establish whether their noise affects the shown measurements.
As before, the noise is quantified by means of the ASD, which corresponds to the square-root of the PSD, describing the spectral distribution of the noise.

\begin{figure}[tbp]
	\centering
	\includegraphics[width=8.5cm]{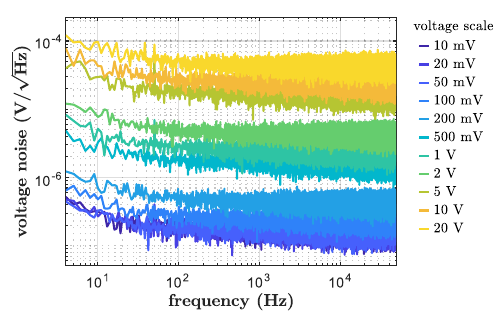}
	\caption{\textbf{Oscilloscope noise.} 
	    Voltage noise of the 5242D PicoScope from Picotech for different voltage scales, as measured for a short-circuited input.
	}
	\label{fig:PicoscopeNoise}
\end{figure}

\begin{figure*}[tbp]
	\centering
	\includegraphics[width=17cm]{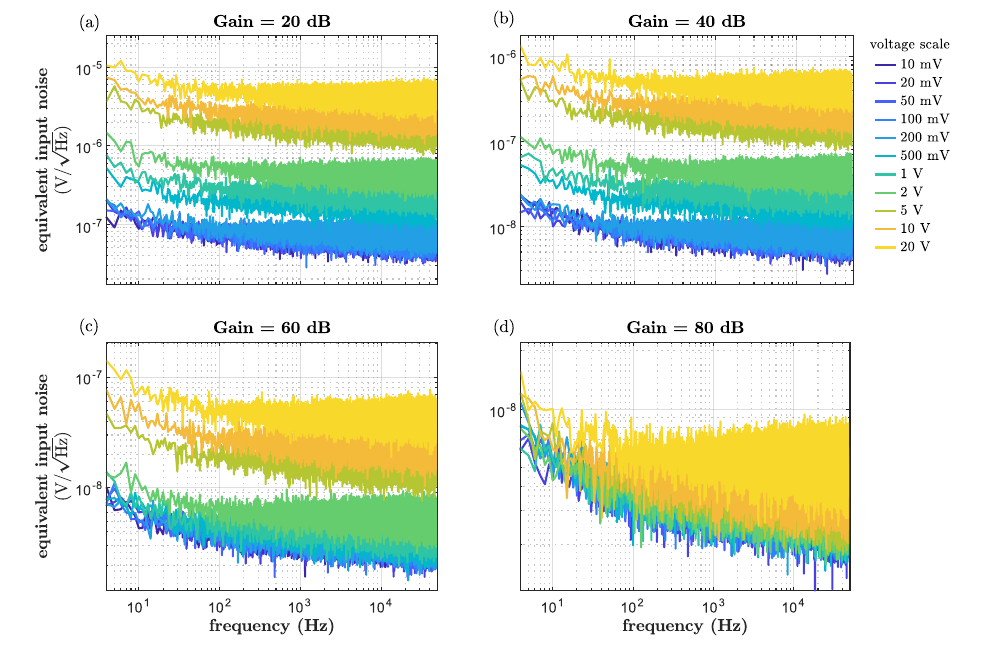}
	\caption{\textbf{Amplifier + Oscilloscope noise.} 
	    Equivalent input noise (EIN) of the Femto voltage amplifier DLPA-100-BD for different gains, each at different voltage scales at the oscilloscope, as measured for a short-circuited input of the amplifier.
	}
	\label{fig:FemtoNoise}
\end{figure*}

\begin{figure}[tbp]
	\centering
	\includegraphics[width=8.5cm]{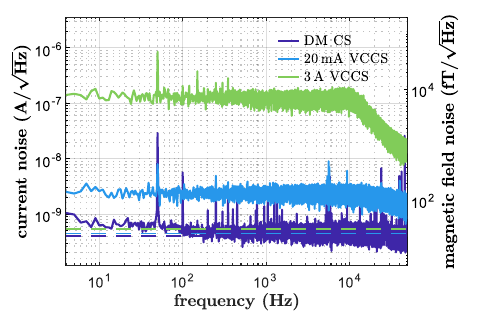}
	\caption{\textbf{Current sources.} 
	    Current noise of the low-noise current source from DM Technologies and the two VCCSs from Tab.~\ref{tab:CS}. The horizontal dashed lines correspond to the noise floor of the amplifier, converted into current noise, operating as in the measurements for the different sources. The measured noise level for the low-noise current source is limited by the noise of the measuring device.
	}
	\label{fig:CSsNoise}
\end{figure}

\begin{figure}[tbp]
	\centering
	\includegraphics[width=8.5cm]{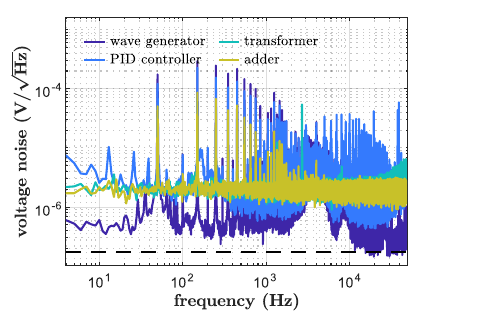}
	\caption{\textbf{Other devices.} 
	    Voltage noise of the function generator from Agilent, the PID controller from Stanford Research System, the transformer from Omicron Lab and the adder used to add artificial noise to the control circuit. The horizontal dashed lines correspond to the noise floor of the oscilloscope with the voltage scale used for the measurements.
	}
	\label{fig:DevicesNoise}
\end{figure}

The oscilloscope noise is estimated by short-circuiting the input and calculating the ASD of the voltage signal for every voltage range of the device, obtaining the spectra shown in Fig.~\ref{fig:PicoscopeNoise}.
For the amplifier, we short-circuit pins 1, 2 and 3, corresponding to the non-inverting input, the inverting input and the grounding, and connect the output to the oscilloscope. From the raw data collected, we calculate the Equivalent Input Noise (EIN), which corresponds to the output noise divided by the gain of the amplifier, at every possible gain and for every voltage scale of the oscilloscope. The results are shown in Fig.~\ref{fig:FemtoNoise} for amplifications of 20, 40, 60, and $80\,$dB.
Both the measurements for the oscilloscope and the amplifier show that the noise level increases with an increasing voltage range, due to a decrease in the sensitivity of the oscilloscope for a fixed bit depth. At the same time, the EIN of the amplifier decreases for increasing gain. 
Therefore, in all noise measurements throughout this work, we set the smallest voltage scale compatible with the chosen amplification, in order to maximize the sensitivity of the measuring devices to the input noise.
The horizontal lines in the corresponding plots represent the noise floor of the amplifier EIN, as measured with the same gain and voltage scale. We estimate this by integrating the relative spectrum for all frequencies above 1$\,$kHz. If the noise of the signal of interest lies above this line, we state that the measurement was not affected by the measuring devices.

Figure \ref{fig:CSsNoise} shows the noise of the three current sources used in this study: the low-noise current source from DM-Technologies (DMT) and the two VCCSs. For the measurements, the output of the DMT source is set to 0, while the control analog input of the VCCSs is short-circuited. In each case, a high precision \unit{10}{\ohm} resistor is connected at the output and the voltage drop is amplified by the voltage amplifier, whose output is connected to the oscilloscope. The detected voltage signal is converted back into a current taking into account the amplifier gain and the resistor, so that the current noise can be calculated.
An estimation of the noise floor for different devices is obtained by integrating the measured spectra for all frequencies up to $10\,$kHz, resulting in $7\times10^{-10}\,$A/$\sqrt{\text{Hz}}$,  $3.3\times10^{-9}\,$A/$\sqrt{\text{Hz}}$ and $2\times10^{-7}\,$A/$\sqrt{\text{Hz}}$ for the DMT source, the $20\,$mA source (VCCS1) and the $3\,$A source (VCCS2), respectively. Notice that the DMT source spectrum lies on the horizontal line indicating the EIN level of the amplifier. This means that the measurement was limited by the device and that the actual current noise is lower (specified at 10$\,$pA/$\sqrt{\text{Hz}}$). The noise floor of the $3\,$A VCCS2 is well above the estimated sensitivity of our setup ($2.46\,\times10^{-9}\,$A/$\sqrt{\text{Hz}}$, cf.~Sec.~\ref{subsec:results_characterization}), making it possible to stabilize the output current.

Figure \ref{fig:DevicesNoise} shows the measured noise spectra for the waveform generator from Agilent (outputting a DC-signal at null level), the PID controller from Stanford Research System (setpoint set internally to 0 and $P=1$, $I=D=0$), the transformer from Omicron Lab and the adder used to add artificial noise (both with short-circuited inputs). All the devices are directly connected to the oscilloscope.


\newpage

\end{document}